 \documentclass[journal,12pt, onecolumn,draftclsnofoot,]{IEEEtran}
\usepackage{amssymb}
\usepackage{amsmath}

\usepackage{array}

\usepackage[dvipsnames]{xcolor}	% load before tikz!!!

\usepackage{pgfplots}
\usepackage{algorithm, algpseudocode}
\usepackage{enumerate}
\usepackage{graphicx}
\usepackage{bm}
\usepackage{threeparttable}
\usepackage{pgfplotstable}
\pgfplotsset{compat=newest}
\pgfplotsset{plot coordinates/math parser=false}

\usepackage[nospace]{cite}
\usepackage{subcaption}
\usepackage{footnote}
\makesavenoteenv{table}
\usetikzlibrary{quotes,arrows.meta}
\usetikzlibrary{positioning}
\usetikzlibrary{spy}
\usetikzlibrary{backgrounds}
\usetikzlibrary{3d} %for including external image 

\definecolor{myred}{RGB}{161,23,23}
\definecolor{myblue}{RGB}{23,32,161}
\definecolor{mygreen}{RGB}{66, 172,21}
\definecolor{myorange}{RGB}{245,146,33}
\definecolor{mykaki}{RGB}{209,215,38}
\definecolor{myviolet}{RGB}{205,99,243}

\graphicspath{{./figs/}}

\begin{document}
%
% paper title
\title{Efficient Majority Voting in Digital Hardware}
\author{Stefan~Baumgartner, Mario~Huemer~\IEEEmembership{Senior Member,~IEEE}, and Michael~Lunglmayr,~\IEEEmembership{Member,~IEEE}%
\thanks{S. Baumgartner and M. Huemer are with the JKU LIT SAL eSPML Lab, Johannes Kepler University Linz, Austria, and with the Institute of Signal Processing, Johannes Kepler University Linz, Austria.}%
\thanks{M. Lunglmayr is with the Institute of Signal Processing, Johannes Kepler University Linz, Austria.}%
\thanks{e-mails: \{stefan.baumgartner, mario.huemer, michael.lunglmayr\}@jku.at}}% <-this % stops a space
%
%
% author names and IEEE memberships

% make the title area
\maketitle
\begin{abstract}
In recent years, machine learning methods became increasingly important for a manifold number of applications. However, they often suffer from high computational requirements impairing their efficient use in real-time systems, even when employing dedicated hardware accelerators.
Ensemble learning methods are especially suitable for hardware acceleration since they can be constructed from individual learners of low complexity and thus offer large parallelization potential. For classification, the outputs of these learners are typically combined by majority voting, which often represents the bottleneck of a hardware accelerator for ensemble inference.
In this work, we present a novel architecture that allows obtaining a majority decision in a number of clock cycles that is logarithmic in the number of inputs. We show, that for the example application of handwritten digit recognition a random forest processing engine employing this majority decision architecture implemented on an FPGA allows the classification of more than seven million images per second.
\end{abstract}


\section{Introduction}
Hardware accelerators for machine learning algorithms have become an important research topic in recent years. For example, a large amount of research has been spent on accelerators for neural networks, e.g. \cite{FPGASVM2010,CNNFPGA,TCASSPIKEMNIST21,Medus19, NNplus1, NNplus2, NNplus3, NNplus4}. Alternatives to neural networks seem to have been investigated to a lesser extent. However, as it has been thoroughly demonstrated, e.g. in \cite{dowennedhundredsofclass}, alternative methods such as ensemble learning methods (e.g. random forests) can have a comparable inference performance to neural networks. From a hardware implementation perspective, members of the group of ensemble learning methods often have the advantage that they rely on fundamental operations other than the scalar product (as neural networks do), that can be implemented with less complexity in digital hardware. Furthermore, such methods often also show a high parallelization potential.

This work is structured as follows. In Sect.~\ref{sect:EnsembleLearning}, we discuss ensemble learning and one of its subgroups, namely random forests, that are used in this work. We advocate a simple architecture suitable for parallel processing of multiple trees in digital hardware. We then present a novel architecture for majority voting in Sect.~\ref{sec:Majority_Decision}. Combining these building blocks enables an efficient random forest acceleration, by exploiting parallelism within its structure. We derive equations for the required number of clock cycles to make a majority decision as well as to fully process a random forest. In Sect.~\ref{sec:Results}, we show FPGA synthesis results demonstrating the low area requirements and high clock speeds that are achievable with this architecture. Finally, we present synthesis and accuracy results for a full random forest engine incorporating the majority decision block for the example of handwritten digit recognition on the MNIST database~\cite{LeCun10}. These results demonstrate the high inference speed that is achievable with this architecture as, for this example, we were able to obtain a $96\,\%$ recognition rate with a processing speed that allows classifying over 7~million $28 \times 28$ pixel images per second.

\section{Ensemble Learning}
\label{sect:EnsembleLearning}
Ensemble learning uses multiple learners (the ensemble) whose outputs are combined to make a final decision. An ensemble is comprised of so-called {\em base learners} \cite{freund}. For classification algorithms, the learners' outputs are typically combined by a majority decision. 
Famous techniques for ensemble learning methods are bootstrap aggregation (also called bagging), which is used for so-called random forests \cite{breiman1984}, or boosting \cite{freund}. From a hardware implementation perspective, using ensemble learning methods has a variety of advantages. Since multiple learners, typically operating independently of each other on the data, are employed, such methods have a high parallelization potential. The computationally more expensive inference operations of the learners can be calculated in parallel if the available hardware resources allow. The final bottleneck of the whole inference task is then only the combination of the learners' outputs. In the following sections, we describe and analyze an efficient architecture combining the learners' outputs by a majority vote.

\subsection{Decision Tree Learning}
A decision tree consists of levels of decision nodes and a final (leaf) level where the inference result is output. We will consider only axis parallel binary trees since they are mainly utilized for random forests. 
For inference, a tree is applied to an input vector $\mathbf{x}$. Starting from its root node, a decision tree is processed level by level, moving along its edges depending on the results obtained from the decision nodes~\cite{tesl}. 
Each comparison node compares one coordinate of a $p \times 1$ input vector $\mathbf{x}$ with the comparison value of the current node. As this results in splitting an area of $\mathbb{R}^p$ into two parts, we will call the coordinate of a node \emph{splitting coordinate} and the comparison value \emph{splitting value}. In this work, the splitting coordinates and the splitting values are learned from training data by the CART~\cite{breiman1984} algorithm for random forests. 

\subsection{State-of-the-Art on Tree Inference Hardware Accelerators}
In \cite{ImplDecTrees, PipelinedDecTree}, implementations of tree inference structures on FPGAs have been discussed. In these works, the main focus was on building pipelined versions of tree accelerators relying on parallelized node implementations in logic. There, the described versions range from accelerators using single node hardware up to implementing all nodes in parallel. In the latter, the parallelizing node comparisons and combining  comparison results require a large amount of distributed logic and distributed memory resources.

For random forest implementations, one has to process multiple trees to obtain a classification result. This opens the possibility of processing multiple trees in parallel, which, according to the authors' opinion, is preferable to parallelizing the processing of a single tree. For this, we advocate the use of a lightweight tree inference architecture, as discussed in the next section. Although the structure is similar to the 
``Universal Node Architecture'' of \cite{ImplDecTrees}, it is even more simplified and it utilizes an FPGA's Block RAMs for implementation.

For random forests, the classifications of the individual trees have to be combined, typically using a (unweighted) majority decision. As this represents a bottleneck, an efficient hardware acceleration of this task is crucial. An interesting approach to obtain a majority vote in hardware is described in~\cite{Barbareschi15, Barbareschi21}. There, multiple sorting networks are utilized to accomplish the majority decision. However, the number of clock cycles needed for sorting with the proposed sorting networks scales linearly with the number of trees $T$ in a random forest as well as with the number of classes $K$: the complexity in terms of number of clock cycles is of $\mathcal{O}(T+K)$. Even with more time-optimized sorting networks (having a more complicated structure), the number of clock cycles would still be of $\mathcal{O}((\log(T))^2)$~\cite{IntroToAlgorithms}. Our majority decision block, which is detailed in the following, has a clock cycle complexity of $\mathcal{O}(\log(T))$, which is clearly beneficial for a large number of decision trees.

\subsection{Accelerating Inference for Random Forests}
\begin{figure}[ht]
	\centering
	\includegraphics[width=.6\columnwidth]{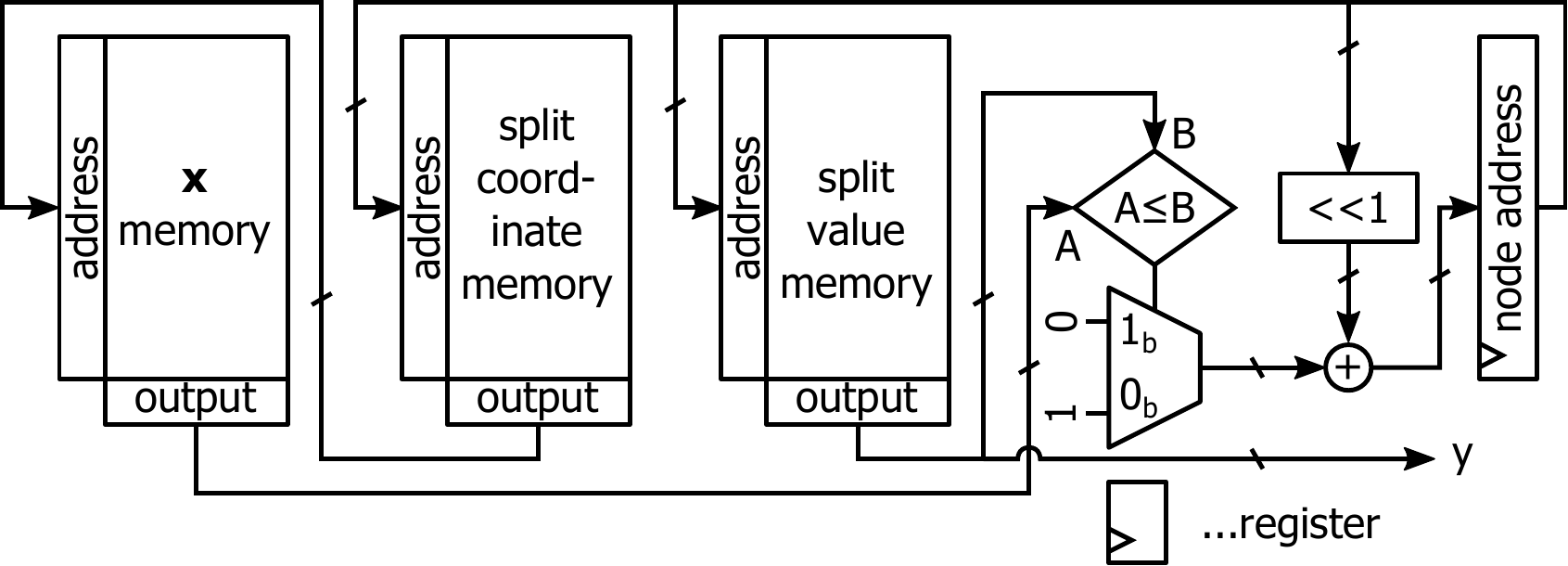}
	\caption{Tree inference architecture.}
	\label{fig:TreeProcessor}
\end{figure}

When using an appropriate numbering of the nodes, starting at the root node with `$1$', the transition from a node of one level of a tree to a node of a lower level can be implemented very efficiently by a left shift of the (binary) node number followed by an increment of one if the node comparison was true, or by no increment otherwise.
This leads to the architecture for tree inference that is schematically shown in Fig.~\ref{fig:TreeProcessor}. It consists of two memories representing the tree (implemented as Block RAMs on the FPGA), the split coordinate memory, and the split value memory. The latter also contains the values of the leaf nodes that are output as classification result $y$ if the leaf level is reached. The addresses of the memory entries correspond to the node number of the trees. The simple node address calculation hardware is depicted at the bottom of Fig.~\ref{fig:TreeProcessor}. Due to the constant left shift when moving down the levels of the tree, the addition of one can be implemented by simply setting the least significant bit of the node address. This structure allows processing one level of a tree within three clock cycles (one clock cycle to get the outputs of the tree memories, one clock cycle to access the corresponding coordinate of $\mathbf{x}$, and one clock cycle to perform the comparison and update the node address). Although the architecture shown in Fig.~\ref{fig:TreeProcessor} is not directly pipelineable, the simplicity of this structure easily allows parallel processing of multiple trees in hardware (the number of trees that can be implemented in parallel is mainly defined by the number of available memories), which is beneficial for a random forest scenario. This is also supported by the fact that in a random forest typically the trees are different from each other. 
Thus, for a pipelined implementation one might have to exchange the tree parameters (the comparison values and the coordinates) during inference, complicating the overall structure and potentially emptying the pipeline. The proposed structure, however, easily allows implementing $40$ trees of height $14$ in parallel plus the majority decision architecture on state-of-the-art FPGAs, as we demonstrate in Sect.~\ref{ssec:Application_Example_Handwritten_Digit_Recognition}.

\section{Majority Decision}
\label{sec:Majority_Decision}
The majority decision part of a random forest maps a vector of $T$ integer numbers (the class outputs of the individual trees) to the number occurring most frequently.
In Fig.~\ref{fig:Architecture_Majority_Dec} we schematically show the proposed architecture for performing a majority vote in hardware. As described below, the complexity scales only logarithmically with $T$. 
 However, as the drawing in Fig.~\ref{fig:Architecture_Majority_Dec} shows, the area requirements for the proposed architecture scale linearly with $K$. We will first describe the iterative version of the architecture (as it is drawn) and then detail how the architecture can be pipelined.
\subsection{Architecture}
\label{ssec:Architecture}
\subsubsection{Iterative Architecture}
\label{sssec:Iterative_Architecture}
\begin{figure}[ht]
	\centering
	\includegraphics[width=.4\columnwidth]{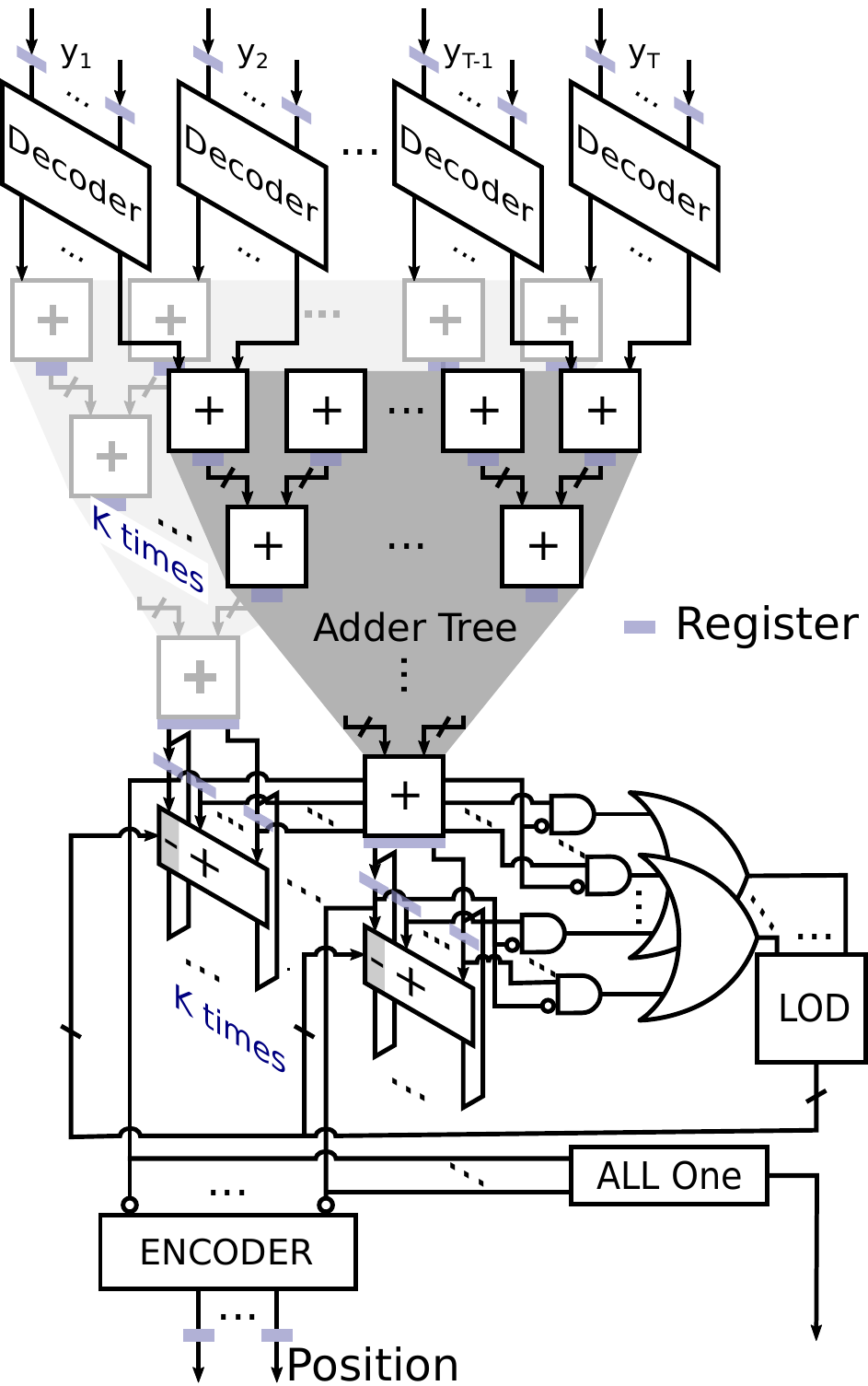}
	\caption{Architecture of the majority decision block.}
	\label{fig:Architecture_Majority_Dec}
\end{figure}
In a first step, the $T$ integer input class values $y_i \in {0, ..., K-1}, i=0, ..., T-1$ are decoded into their one-hot representation, i.e. into vectors of length $K$ containing all `$0$'s but a `$1$' at the $y_i$-th position. Then, the count of the members of each class in the set of input values should be determined. Hence, all bits at the $j$-th position, $j = 0, ..., K-1$, of the one-hot vectors have to be summed up. For this purpose, $K$ parallel adder trees are employed.  In order to obtain an efficient design, the adders of every adder tree operate with the minimum required input bit width, which is one in the first stage and increases by one in each subsequent stage. That is, the first stage of every adder tree consists of $\left\lceil\frac{T}{2}\right\rceil$ one-bit adders with an output bit width of two, while at its last stage one $\lceil\log_2(T)\rceil$-bit adder is used. To maintain a high clock frequency, registers (drawn in violet in Fig.~\ref{fig:Architecture_Majority_Dec}) are introduced after every stage of an adder tree. Every output of the $K$ adder trees is the count of the corresponding class in the set of input class values. For further processing, the class counts have to be represented in the two's complement and thus a \lq$0$\rq\,sign bit is appended in front of every output of the adder trees (at the position over the ``-'' of the subtractors; the left inputs are the subtrahends). 

These class counts in two's complement serve as the initial values of the registers feeding the $K$ parallel subtractors, shown below the adder trees in Fig.~\ref{fig:Architecture_Majority_Dec}. 
The residual part of the circuit shown in Fig.~\ref{fig:Architecture_Majority_Dec} is used to iteratively subtract the most significant `$1$'~bit of all class counts (determined by the OR gates and the leading one detector - LOD) from all individual count sums. For this, only those counts that are still positive will be considered (AND gates with negated inputs from the sign bits). This will be done until all class counts are negative. The index of the last non-negative class count\footnote{In case of a draw the class with the highest class number is the output class.} identifies the class that was output by the learners with maximum frequency. This number is output by the encoder at the bottom of Fig.~\ref{fig:Architecture_Majority_Dec}. As one is interested in the class number of the cycle before all class counts are negative, the encoder's output is delayed by a register stage.

\subsubsection{Pipelined Architecture}
\label{sssec:Pipelined_Architecture}
The architecture described in the last section works in an iterative manner, since the adder tree output values are decreased iteratively until all values are negative, which might take up to $\lceil \log_2(T)\rceil + 1$ iterations. Thus, a majority decision cannot be made at every clock cycle. In many application scenarios, this is not a problem, since not at every clock cycle classification results from e.g. decision tree classifiers of a random forest are available, which is also the case for our random forest implementation. However, in some scenarios, a pipelined majority decision architecture might be needed and thus we make slight adaptions such that the proposed architecture can provide a majority decision at every clock cycle. That is, the iterative subtraction of the adder tree output values by the LOD output values is ``unfolded'' to $\lceil \log_2(T)\rceil$ stages. In each stage, one LOD decision and $K$ parallel subtractions take place. If all sign bits in a stage are \lq$1$\rq, the outputs of the previous stage are passed to all following stages, discarding their outputs.

\subsection{Analysis}
In the following, we consider the number of clock cycles that are needed to make a majority decision with the architectures described in Sect.~\ref{ssec:Architecture}. Since a decoder is a low-complexity block, no registers between decoders and adder trees are introduced. The $K$ parallel adder trees consist of $\lceil\log_2(T)\rceil$ adder stages, and thus $\lceil\log_2(T)\rceil$ clock cycles after the input valid strobe the class counts are available at the outputs of the adder trees. For the iterative architecture described in Sect.~\ref{sssec:Iterative_Architecture}, the required number of clock cycles can be determined as follows. The adder tree results are stored in registers before the iterative subtraction of the class counts starts, which adds another clock cycle. Given the stored class counts in the registers, the residual number of clock cycles is equal to the number of `$1$'~bits in the bit pattern of the maximum class count's value plus one cycle to make all count values negative.
The best case is, when the maximum count is a power of two, giving a lower limit of the overall required number of clock cycles for the majority decision
\begin{gather}
N_{iter,min} = \lceil\log_2(T)\rceil + 3\,.
\end{gather}
The worst case occurs, when the binary maximum class count contains only \lq$1$\rq\,bits at bit positions behind the most significant \lq$1$\rq\,bit (i.e. the maximum class count is a power of two minus one). As the maximum number of \lq$1$\rq\,bits is $\lfloor\log_2(T)\rfloor$, the upper bound of the overall required number of clock cycles for the majority decision is
\begin{gather}
N_{iter,max} = \lceil\log_2(T)\rceil + \lfloor\log_2(T)\rfloor + 2\,.
\end{gather}

Since only a part of the whole architecture works iteratively, there is no need to wait until the majority decision has finished to feed the next input values into the architecture. This means, every $\lceil\log_2(T)\rceil + 1$ clock cycles a new majority decision can be started, which guarantees that the proposed iterative architecture works properly for all constellations of input values.  

For the pipelined version of the majority decision architecture detailed in Sect.~\ref{sssec:Pipelined_Architecture}, the number of needed clock cycles is constant:
\begin{gather}
N_{pipe} = \lceil\log_2(T)\rceil + \lfloor\log_2(T)\rfloor+1\,.
\end{gather}
Here, the worst case for the maximum class count has to be assumed, but the iterative subtraction of the class counts can in this case already be abandoned when all values but one are negative. However, this is only an initial delay and afterward a new majority decision is available at every clock cycle.

\section{Results}
\label{sec:Results}
\subsection{Synthesis Results for the Majority Decision Block}
\begin{figure}[t]
	\begin{subfigure}{\columnwidth}
		\centering
		\begin{tikzpicture}
		\begin{axis}[compat=newest, width=8.2cm, height=4cm, grid=both, ylabel={\scriptsize $f_{max}$ in MHz}, ymax = 750, ymin = 0, xmax=512, xmin=0, xlabel={\scriptsize $T$}, scaled ticks = false,  x tick label style={/pgf/number format/.cd,fixed,precision=2,/tikz/.cd}, ticklabel style={font=\scriptsize}, legend style={at={(0.5,1.02)}, anchor=south}, ytick={0,100,200,300,400,500,600,700}, legend columns = {3}, legend cell align=left, legend style={font=\scriptsize}, every axis plot/.append style={thick}]
		\pgfplotstableread[col sep=semicolon]{f_max_Major_Dec.csv}\datatable;
		\pgfplotstabletranspose[header=true, colnames from=N_CATEGORIES/N_INPUTS, input colnames to=n_inputs]\transposeddatatable{\datatable};
		
		\addplot[color=myblue, mark=o, smooth] table[col sep=semicolon, x = n_inputs, y = 2]{\transposeddatatable};
		\addlegendentry{\scriptsize $K = 2$};
		\addplot[color=mygreen, mark=o, smooth] table[col sep=semicolon, x = n_inputs, y = 5]{\transposeddatatable};
		\addlegendentry{\scriptsize $K = 5$};
		\addplot[color=myred, mark=o, smooth] table[col sep=semicolon, x = n_inputs, y = 15]{\transposeddatatable};
		\addlegendentry{\scriptsize $K = 15$};
		%\addplot[color=myorange, mark=o, smooth] table[col sep=semicolon, x = n_inputs, y = 30]{\transposeddatatable};
		%\addlegendentry{\scriptsize $K = 30$};
		\addplot[color=myviolet, mark=o, smooth] table[col sep=semicolon, x = n_inputs, y = 100]{\transposeddatatable};
		\addlegendentry{\scriptsize $K = 100$};
		%\addplot[color=mykaki, mark=o, smooth] table[col sep=semicolon, x = n_inputs, y = 200]{\transposeddatatable};
		%\addlegendentry{\scriptsize $K = 200$};
		\addplot[color=black, mark=o, smooth] table[col sep=semicolon, x = n_inputs, y = 500]{\transposeddatatable};
		\addlegendentry{\scriptsize $K = 500$};
		\end{axis}
		\end{tikzpicture}
	\end{subfigure}
	\begin{subfigure}{\columnwidth}
		\centering
		\begin{tikzpicture}
		\begin{semilogyaxis}[width=8.2cm, height=4cm, ylabel = {\scriptsize Nr. ALMs}, xlabel = {\scriptsize $T$}, ymin = 5, ymax = 2e5, xmin=0, xmax=512, grid, scaled ticks = false,  x tick label style={/pgf/number format/.cd,fixed,precision=2,/tikz/.cd}, ticklabel style={font=\scriptsize}, every axis plot/.append style={thick}]
		\pgfplotstableread[col sep=semicolon]{ALMs_Major_Dec.csv}\datatable;
		\pgfplotstabletranspose[header=true, colnames from=N_CATEGORIES/N_INPUTS, input colnames to=n_inputs]\transposeddatatable{\datatable};
		\addplot[color=myblue, mark=o, smooth] table[col sep=semicolon, x = n_inputs, y = 2]{\transposeddatatable};
		\addplot[color=mygreen, mark=o, smooth] table[col sep=semicolon, x = n_inputs, y = 5]{\transposeddatatable};
		\addplot[color=myred, mark=o, smooth] table[col sep=semicolon, x = n_inputs, y = 15]{\transposeddatatable};
		%\addplot[color=myorange, mark=o, smooth] table[col sep=semicolon, x = n_inputs, y = 30]{\transposeddatatable};
		\addplot[color=myviolet, mark=o, smooth] table[col sep=semicolon, x = n_inputs, y = 100]{\transposeddatatable};
		%\addplot[color=mykaki, mark=o, smooth] table[col sep=semicolon, x = n_inputs, y = 200]{\transposeddatatable};
		\addplot[color=black, mark=o, smooth] table[col sep=semicolon, x = n_inputs, y = 500]{\transposeddatatable};
		\end{semilogyaxis}
		\end{tikzpicture}
	\end{subfigure}
	\vspace{-0.2cm}
	\caption{Synthesis results for the iterative architecture.}
	\label{fig:f_max_ALMs_vs_T}
\end{figure}
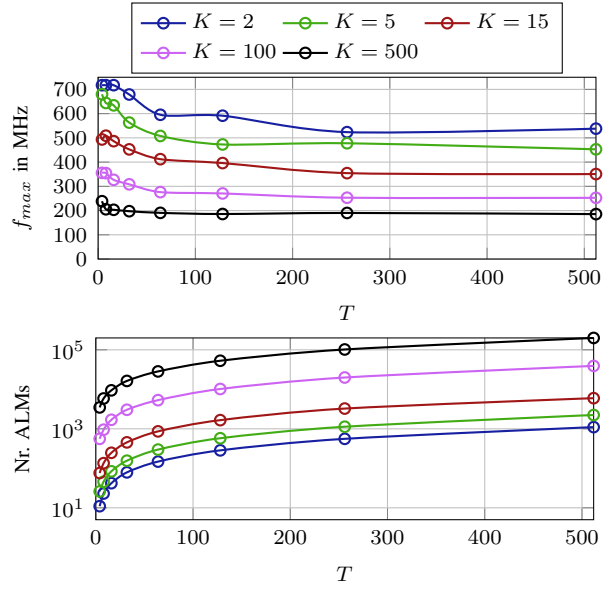

In the following, we present the synthesis results of the proposed majority decision block for an Altera Stratix~V~5SGXEA7 FPGA. To gain insight into how the number of inputs $T$ and classes $K$ influence the restricted maximum clock frequency $f_{max}$ and the number of required adaptive logic modules (ALMs) required, syntheses are conducted for different $T$ and $K$, varying from 4 to 512 and 2 to 500, respectively. The obtained results for the iterative architecture are shown in Fig.~\ref{fig:f_max_ALMs_vs_T}. Especially for more than two classes, it can be observed that $f_{max}$ is only slightly affected by the number of inputs $T$ above around 60 inputs. Obviously, the number of classes $K$ has the larger influence on $f_{max}$. The number of ALMs needed scales linearly with $T$ and $K$ and even a majority decision block for 500 classes and 512 inputs fits into the specified FPGA (199795 ALMs are required for this setup, which corresponds to a logic utilization of 85~\%). 

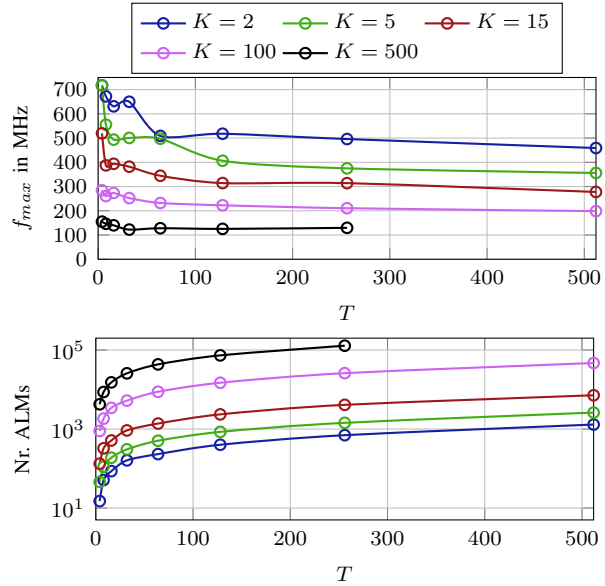
\begin{figure}[t]
	\begin{subfigure}{\columnwidth}
		\centering
		\begin{tikzpicture}
		\begin{axis}[compat=newest, width=8.2cm, height=4cm, grid=both, ylabel={\scriptsize $f_{max}$ in MHz}, ymax = 750, ymin = 0, xmax=512, xmin=0, xlabel={\scriptsize $T$}, scaled ticks = false,  x tick label style={/pgf/number format/.cd,fixed,precision=2,/tikz/.cd}, ticklabel style={font=\scriptsize}, legend style={at={(0.5,1.02)}, anchor=south}, ytick={0,100,200,300,400,500,600,700}, legend columns = {3}, legend cell align=left, legend style={font=\scriptsize}, every axis plot/.append style={thick}]
		\pgfplotstableread[col sep=semicolon]{f_max_Major_Dec_pipelined.csv}\datatable;
		\pgfplotstabletranspose[header=true, colnames from=N_CATEGORIES/N_INPUTS, input colnames to=n_inputs]\transposeddatatable{\datatable};
		
		\addplot[color=myblue, mark=o, smooth] table[col sep=semicolon, x = n_inputs, y = 2]{\transposeddatatable};
		\addlegendentry{\scriptsize $K = 2$};
		\addplot[color=mygreen, mark=o, smooth] table[col sep=semicolon, x = n_inputs, y = 5]{\transposeddatatable};
		\addlegendentry{\scriptsize $K = 5$};
		\addplot[color=myred, mark=o, smooth] table[col sep=semicolon, x = n_inputs, y = 15]{\transposeddatatable};
		\addlegendentry{\scriptsize $K = 15$};
		%\addplot[color=myorange, mark=o, smooth] table[col sep=semicolon, x = n_inputs, y = 30]{\transposeddatatable};
		%\addlegendentry{\scriptsize $K = 30$};
		\addplot[color=myviolet, mark=o, smooth] table[col sep=semicolon, x = n_inputs, y = 100]{\transposeddatatable};
		\addlegendentry{\scriptsize $K = 100$};
		%\addplot[color=mykaki, mark=o, smooth] table[col sep=semicolon, x = n_inputs, y = 200]{\transposeddatatable};
		%\addlegendentry{\scriptsize $K = 200$};
		\addplot[color=black, mark=o, smooth] table[col sep=semicolon, x = n_inputs, y = 500]{\transposeddatatable};
		\addlegendentry{\scriptsize $K = 500$};
		\end{axis}
		\end{tikzpicture}
	\end{subfigure}
	\begin{subfigure}{\columnwidth}
		\centering
		\begin{tikzpicture}
		\begin{semilogyaxis}[width=8.2cm, height=4cm, ylabel = {\scriptsize Nr. ALMs}, xlabel = {\scriptsize $T$}, ymin = 5, ymax = 2e5, xmin=0, xmax=512, grid, scaled ticks = false,  x tick label style={/pgf/number format/.cd,fixed,precision=2,/tikz/.cd}, ticklabel style={font=\scriptsize}, every axis plot/.append style={thick}]
		\pgfplotstableread[col sep=semicolon]{ALMs_Major_Dec_pipelined.csv}\datatable;
		\pgfplotstabletranspose[header=true, colnames from=N_CATEGORIES/N_INPUTS, input colnames to=n_inputs]\transposeddatatable{\datatable};
		\addplot[color=myblue, mark=o, smooth] table[col sep=semicolon, x = n_inputs, y = 2]{\transposeddatatable};
		\addplot[color=mygreen, mark=o, smooth] table[col sep=semicolon, x = n_inputs, y = 5]{\transposeddatatable};
		\addplot[color=myred, mark=o, smooth] table[col sep=semicolon, x = n_inputs, y = 15]{\transposeddatatable};
		%\addplot[color=myorange, mark=o, smooth] table[col sep=semicolon, x = n_inputs, y = 30]{\transposeddatatable};
		\addplot[color=myviolet, mark=o, smooth] table[col sep=semicolon, x = n_inputs, y = 100]{\transposeddatatable};
		%\addplot[color=mykaki, mark=o, smooth] table[col sep=semicolon, x = n_inputs, y = 200]{\transposeddatatable};
		\addplot[color=black, mark=o, smooth] table[col sep=semicolon, x = n_inputs, y = 500]{\transposeddatatable};
		\end{semilogyaxis}
		\end{tikzpicture}
	\end{subfigure}
	\vspace{-0.2cm}
	\caption{Synthesis results for the pipelined architecture.}
	\label{fig:f_max_ALMs_vs_T_pipelined}
\end{figure}

The synthesis results for the pipelined architecture are plotted in Fig.~\ref{fig:f_max_ALMs_vs_T_pipelined}. It can be observed that the behavior of the obtained curves is similar to that of the iterative architecture. However, the maximum clock frequency of the pipelined architecture for a specific setup (fixed $T$ and $K$) is significantly lower than the corresponding one of the iterative architecture. Furthermore, the number of ALMs required for the pipelined architecture is higher and for $T = 512$ inputs and $K = 500$ classes, the design no longer fits into the FPGA.

\subsection{Application Example: Handwritten Digit Recognition}
\label{ssec:Application_Example_Handwritten_Digit_Recognition}
In this section, we describe results for a random forest processing engine comprised of multiple instances of the tree processing units as shown in Fig.~\ref{fig:TreeProcessor} and an instance of the majority decision unit. Processing of a tree with $l$ levels of decision nodes followed by one level of terminal nodes requires $3l+1$ clock cycles (three clock cycles per level for the processing of a decision node and one clock cycle to output the value of the terminal node as $y$). Assuming that $T$ trees can be processed in parallel in digital hardware, and combined with the iterative majority decision architecture, this leads to a worst case number of clock cycles of
\begin{align}
3l+ \lceil\log_2(T)\rceil + \lfloor\log_2(T)\rfloor +3
\label{eqn:allclcks}
\end{align}
to finish the classification of an input vector $\mathbf{x}$ using a random forest of $T$ trees. 

We synthesized the described architecture for a random forest trained on the MNIST handwritten digit database \cite{LeCun10}. For this, we used $40$ trees and 
$14$ levels of decision nodes per tree.
Each tree of the random forest has been learned with a random selection of $75\,\%$ of the $60000$ training images. To obtain the splitting coordinates, only $\sqrt{784}=28$ of the $784$ coordinates (again selected randomly) of each image ($28 \times 28$ pixels) have been allowed to be selected to learn the best splitting coordinates, as it is typical for random forests \cite{tesl}. This allowed obtaining a classification performance on the MNIST test set ($10000$ images) of $96\,\%$ correctly classified digits. For reproducibility, we uploaded the contents of the tree memories to \cite{gitRepoMatlabRF}. Although this performance is below the best classification performances described in the literature \cite{LeCun10}, it is comparable to the state-of-the-art of hardware architectures for this problem in terms of inference accuracy. When calculating the required number of clock cycles using \eqref{eqn:allclcks} for this use case one obtains $3 \cdot 14 + \lceil \log_2(40) \rceil + \lfloor \log_2(40) \rfloor +3 = 56$ clock cycles for the classification of a single image. However, assuming that processing a tree requires more clock cycles than the majority decision, one can input new data vectors $\mathbf{x}$ already after tree processing is finished. That is, after the 56 clock cycles for the first classification one can obtain new classifications every $3\cdot 14 + 1 = 43$ clock cycles\footnote{The specified timing is valid only for the described architecture not including the time required for the data handling.}.
As one can see from Tab.~\ref{tab:MNISTRes}, besides the number of block RAM bits, the hardware requirements of the random forest architecture using the proposed majority decision block are very low. The block RAM bits are used to store the vectors $\mathbf{x}$ as well as the split coordinates and the split values for each tree. The memory is typically also the limiting factor for this architecture for larger problem sizes.
The authors would like to point out that the main aim was not to achieve the best classification performance, but to demonstrate the inference speed that is achievable with this architecture. For the described problem, the architecture is able to classify more than seven~million images per second while still maintaining a, as we think, decent classification performance. Furthermore, we want to point out that the used random forest was applied directly on the image vectors without using any pre-processing at all. This demonstrates the potential of the proposed random forest architecture. A comparison with state-of-the-art implementations of hardware accelerators for the MNIST application is shown in Tab.~\ref{tab:Comparison}. Although some of the compared methods show a better classification performance for the problem at hand, the presented architecture is about $30$ times faster than the fastest state-of-the-art implementation used for comparison.

\begin{table}[t]
	\centering
	\caption{ Synthesis results for the full RF architecture \label{tab:MNISTRes}}
	\setlength\tabcolsep{3pt} % default value: 6pt
	\setlength\extrarowheight{3pt}
	\begin{tabular}{|>{\raggedright}p{2.5cm}|>{\raggedright}p{2.5cm}|>{\raggedright}p{2.5cm}|>{\raggedright}p{2.8cm}|>{\raggedright\arraybackslash}p{2.8cm}|}
		\hline
	\multicolumn{5}{|c|}{FPGA: Stratix~V 5SGXEA7} \\
	\hline
Slices/ALMS & Registers  & DSP blocks & Block RAM in Mbits & Max.~clock in MHz \\
		\hline
$2709$								& $2495$ &$0$&$28.03$  &$303.4$\\
		\hline
\end{tabular}
\end{table}

\begin{table}[t]
	\centering
	\caption{ Literature reports on machine learning hardware for MNIST \label{tab:Comparison}}
	\setlength\tabcolsep{3pt} % default value: 6pt
	\setlength\extrarowheight{3pt}
	\begin{tabular}{|>{\raggedright}m{3cm}|>{\raggedright}m{1.5cm}|>{\raggedright}m{1.5cm}|>{\raggedright}m{1.5cm}|>{\raggedright}m{1.5cm}|>{\raggedright\arraybackslash}m{1.5cm}|}
		\hline
		\centering Work        & \centering\cite{FPGASVM2010}			& \centering\cite{CNNFPGA}	& \centering\cite{TCASSPIKEMNIST21} & \centering\cite{Medus19} & \centering this work \tabularnewline
		Learning Algorithm		& SVM							& CNN & SNN & FCNN & RF \\
%		\hline
		FPGA/Board                    & Stratix III EP3SE260      & Artix~7 & Virtex~7 VC707   & Virtex~7 VC707 & Stratix~V 5SGXEA7\\
%		\hline
		Max.~clock frequ. (MHz)         &$\leq250$              & $300$   & $\approx 100$ & $490.9$&$303.4$\\
	    Clock cycles            & n.r.\footnote{Not reported.}             & $17400$  & n.r. & $2034$ &$56$\\	
  	    Correctly classified	& $98.96\,\%$				& $98.80\,\%$  &$92.92\,\%$  & $98.60\,\%$& $96.00\,\%$    \\
	    Class. images/second  &$\approx 10\;000$& $\approx 17\;250$   & $\approx 320$& $240\;905$ &$7\;055\;813$\\
		\hline
	\end{tabular}
\end{table}

\section{Conclusion}
We presented a novel hardware implementation for finding a majority decision in digital hardware. 
For this architecture, the number of required clock cycles for a majority vote depends logarithmically on the number of its inputs. We analyzed the number of clock cycles for iterative and pipelined variants of the architecture. Furthermore, we showed synthesis results demonstrating the resource requirements as well as the obtainable clock frequencies for different problem sizes. Finally, we demonstrated the capabilities of our approach for majority voting in hardware in combination with a low complexity tree inference architecture.
We applied the resulting random forest hardware implementation to the MNIST dataset and demonstrated that more than seven million classifications per second are possible on the utilized FPGA.

\end{document}